\documentclass[aps,a4paper,prd,floats,nofootinbib,showkeys]{revtex4}

\usepackage{graphicx}
\usepackage{multirow}

\begin{document}

\title{The high energy neutrino cross-section in the Standard Model and
  its uncertainty}
\author{Amanda Cooper-Sarkar}
\affiliation{Particle Physics, University of Oxford, Keble Road,
             Oxford OX1 3RQ, UK}
\author{Philipp Mertsch, Subir Sarkar}
\affiliation{Rudolf Peierls Centre for Theoretical Physics, 
             University of Oxford, 1 Keble Road, Oxford OX1 3NP, UK\\}
\date{\today}
\bigskip
\begin{abstract}
  Updated predictions are presented for high energy neutrino and
  antineutrino charged and neutral current cross-sections within the
  conventional DGLAP formalism of NLO QCD using modern PDF fits.  PDF
  uncertainties from model assumptions and parametrization bias are
  considered in addition to the experimental uncertainties. Particular
  attention is paid to assumptions and biases which could signal the
  need for extension of the conventional formalism to include effects
  such as $\ln(1/x)$ resummation or non-linear effects of high gluon
  density.
  
\end{abstract}

\keywords{Deep Inelastic Scattering, Neutrino Physics, High Energy
  Cosmic Rays.}
\maketitle

\section{Introduction}

Predictions of neutrino cross-sections at high energies have sizeable
uncertainties which derive from the uncertainties on the parton
distribution functions (PDFs) of the nucleon. In the framework of the
quark-parton model, high energy neutrino deep inelastic scattering
(DIS) accesses large values of $Q^2$, the invariant mass of the
exchanged vector boson, and small values of Bjorken $x$, the fraction
of the momentum of the incoming nucleon taken by the struck
quark. Thus in evaluating uncertainties on high energy neutrino DIS
cross-sections it is important to use the most up-to-date information
from the experiments at HERA, which have accessed the lowest $x$ and
highest $Q^2$ scales to date. H1 and ZEUS have now combined the data
collected in the years 1994--2000 to give very accurate inclusive
cross-sections in the range $ 6 \times 10^{-7} < x < 0.65 $ and $
0.045 < Q^2 < 30000$ GeV$^2$~\cite{Aaron:2009wt}. We consider modern
PDF sets which include these data in order to provide the best
benchmark cross-section for experiments searching for high energy
cosmic neutrinos such as ANITA \cite{Gorham:2008yk}, IceCube
\cite{Abbasi:2010ak}, the Pierre Auger Observatory
\cite{Abraham:2007rj} and RICE \cite{Kravchenko:2011im}, as well as
forthcoming experiments such as ARA \cite{Allison:2011wk}, ARIANNA
\cite{Gerhardt:2010js}, JEM-EUSO \cite{Takahashi:2009zzc} and LUNASKA
\cite{James:2009sf}.

Conventional PDF fits use the next-to-leading-order (NLO) DGLAP
formalism~\cite{Altarelli:1977zs,Gribov:1972ri,Lipatov:1974qm,Dokshitzer:1977sg}
of QCD to make predictions for DIS cross-sections of leptons on
hadrons. At low $x$ it may be necessary to go beyond the DGLAP
formalism in order to sum $\ln(1/x)$ diagrams, as in the BFKL
formalism~\cite{Kuraev:1977fs,Balitsky:1978ic,Lipatov:1985uk} (for
recent work see
Refs.~\cite{Altarelli:2005ni,White:2006xv,Ciafaloni:2006yk,Rojo:2009us,Caola:2009iy}),
or to even consider non-linear terms as in the colour glass condensate
model~\cite{Gelis:2010nm,Goncalves:2010ay}. The present paper is
concerned with estimating the uncertainties on high energy neutrino
cross-sections in the conventional NLO DGLAP formalism. However, the
input assumptions of some of the PDF fits are arguably beyond this
formalism, and such cases will be highlighted.

Pioneering estimates of high energy neutrino cross-sections were
obtained at leading-order~\cite{Gandhi:1998ri} using a PDF set
(CTEQ4m) which is now well out of date, yet these values continue to
be used in analysing current data (e.g. from ANITA
\cite{Gorham:2008yk}). Two of the present authors (CSS)
evaluated~\cite{CooperSarkar:2007cv} the cross-sections at NLO using
the ZEUS PDFs which included more modern HERA data
\cite{Chekanov:2002pv} and using a systematic procedure for estimating
PDF uncertainties. More recently the cross-sections have been
calculated in Ref.~\cite{Connolly:2011vc} (CTW) using the MSTW2008NLO
PDFs, which use HERA data from the same vintage as the ZEUS PDFs and
also account for PDF uncertainties. However, neither of these more
recent works use the newly combined HERA results which have increased
the accuracy of low $x$ data by up to a factor of 3 (see Fig.3 of
Ref.~\cite{Aaron:2009wt}).  It is the purpose of the present paper to
re-evaluate the high energy cross-sections using the most up-to-date
PDF sets, with particular emphasis on those which do use these
precise, combined HERA data. The calculation is made using PDFs which
were evaluated in NLO DGLAP fits, and our calculation of the neutrino
structure functions and cross-sections is also made consistently at
NLO. The input PDFs we consider treat heavy quarks by using
general-mass-variable-flavour number schemes
\cite{Thorne:1997ga,Thorne:2006qt}, and we have also used such a
scheme in our calculation of the structure functions. However, the
difference between the use of a general-mass or a zero-mass scheme in
the latter part of this procedure is negligible since the neutrino
cross-sections are dominated by scattering at high $Q^2$. We consider
carefully the source of uncertainty on the input PDFs; these derive
not only from the experimental uncertainty on the input data but also
from model assumptions, and the form of the PDF parameterization. It
is important to quantify these uncertainties carefully in order to be
able to probe non-perturbative QCD effects at low $x$
\cite{Anchordoqui:2006ta} and/or new physics beyond the Standard Model
\cite{Kusenko:2001gj,Anchordoqui:2005ey,PalomaresRuiz:2005xw,Anchordoqui:2010hq}
through a determination of the DIS cross-section using cosmic
neutrinos which have energies extending up at least to $\sim
10^{11}$~GeV \cite{Anchordoqui:2009nf}.

\section{Formalism}

The kinematics of lepton hadron scattering is described in terms of
the variables $Q^2$, Bjorken $x$, and $y$ which measures the energy
transfer between the lepton and hadron systems.  The double
differential charged current (CC) cross-section for neutrino and
antineutrino production on isoscalar nucleon targets is given by
\cite{Devenish:2004pb}
\begin{equation}
 \frac{\mathrm{d}^2\sigma(\nu (\bar{\nu}) N)}{\mathrm{d}x~\mathrm{d}Q^2} = 
 \frac{G_\mathrm{F}^2 M_W^4}{4\pi(Q^2 + M_W^2)^2 x} 
 \sigma_\mathrm{r}(\nu (\bar{\nu}) N),
\label{eqn:sigma}
\end{equation}
where the reduced cross-sections $\sigma_\mathrm{r}(\nu (\bar{\nu})
N)$ are 
\begin{equation} 
 \sigma_\mathrm{r}(\nu N) = 
 \left[Y_ + F_2^{\nu} (x, Q^2) - y^2 F_\mathrm{L}^{\nu} (x, Q^2) 
 + Y_ - xF_3^{\nu} (x, Q^2) \right],
\label{eqn:nu}
\end{equation}
\begin{equation} 
\sigma_\mathrm{r}(\bar{\nu} N) = \left[Y_ + F_2^{\bar{\nu}}(x, Q^2) - y^2 
 F_\mathrm{L}^{\bar{\nu}}(x, Q^2) - Y_ - xF_3^{\bar{\nu}}(x, Q^2) \right],
\label{eqn:nubar}
\end{equation}
and $F_2$, $xF_3$ and $F_\mathrm{L}$ are related directly to quark
momentum distributions, with $Y_{\pm} = 1 \pm (1-y)^2$.

The QCD predictions for these structure functions are obtained by
solving the DGLAP evolution equations at NLO in the
\mbox{$\overline{\mathrm{MS}}$} scheme with the renormalisation and
factorization scales both chosen to be $Q^2$.  These equations yield
the PDFs at all values of $Q^2$ provided these distributions have been
input as functions of $x$ at some input scale $Q^2_0$.

In QCD at leading order, the structure function $F_\mathrm{L}$ is
identically zero, and the structure functions $F_2$ and $xF_3$ for
charged current neutrino interactions on isoscalar targets can be
identified with quark distributions as follows:
\begin{equation}
 F_2^{\nu} = x (u + d + 2s + 2b + \bar{u} + \bar{d} + 2\bar{c}),\\ 
 \quad
 xF_3^{\nu} = x (u + d + 2s + 2b - \bar{u} - \bar{d} - 2\bar{c}),
\end{equation}
and for antineutrino interactions,
\begin{equation}
 F_2^{\bar{\nu}} = x (u + d + 2c +\bar{u} + \bar{d} + 2\bar{s} + 2\bar{b}),\\ 
 \quad
 xF_3^{\bar{\nu}} = x (u + d + 2c - \bar{u} - \bar{d} - 2\bar{s} - 2\bar{b}).
\end{equation}
At NLO these expressions must be convoluted with appropriate
co-efficient functions in order to obtain the structure functions (and
$F_\mathrm{L}$ is no longer zero) but these expressions still give us
a good idea of the dominant contributions.  The contribution of the
$b$ quark will be suppressed until scales $\sim m_t^2$, since the CKM
element $V_{tb}\sim 1$. Although the dominant contributions to the CC
cross-sections come from $Q^2 \sim M_W^2 \ll m_t^2$, this does not
mean that the $b$ contribution is always suppressed, because the
relevant scale for $t$ production is the virtual boson-nucleon
centre-of-mass energy, $W^2\sim Q^2/x$, and the high energy
cross-sections are dominated by contributions from very small $x \sim
M_W^2/2m_N E_\nu \approx M_W^2 / s$ (see Fig.~\ref{fig:kinreg}). This
point had been overlooked in earlier work, including our
own~\cite{CooperSarkar:2007cv}.

The neutral current cross-sections on isoscalar targets are given by
expressions similar to Eqs.~(\ref{eqn:sigma}, \ref{eqn:nu},
\ref{eqn:nubar}), with the $W$ propagator replaced by the $Z$
propagator, while the leading order expressions for the structure
functions given by
\begin{eqnarray}
  F_2^{\nu,\bar{\nu}} &=& x \left[\frac{(a_u^2+v_u^2 + a_d^2+v_d^2) }{2} 
  (u + \bar{u} + d +\bar{d}) + (a_d^2+v_d^2)(s+b+\bar{s}+\bar{b}) 
+ (a_u^2+v_u^2)(c + \bar{c})\right],\\ \nonumber
   xF_3^{\nu,\bar{\nu}} &=& x [(u -\bar{u} + d-\bar{d})(v_u a_u +v_d a_d)],
\end{eqnarray}
where $v_u$, $v_d$, $a_u$, $a_d$ are the neutral current vector and
axial-vector couplings for $u-$ and $d-$type quarks.

\section{Parton Density Functions}

The PDF4LHC group has recently benchmarked modern parton density
functions~\cite{Alekhin:2011sk}.  Since our concern is with high
energy neutrino cross-sections, rather than with LHC physics, we focus
on PDF sets which make use of the newly combined accurate HERA data
\cite{Aaron:2009wt}. Of all the PDFs considered by the PDF4LHC only
HERAPDF1.0 \cite{Aaron:2009wt}) and NNPDF2.0 \cite{Ball:2010de} used
these data. However there has been a subsequent update of the CTEQ6.6
\cite{Tung:2006tb} PDFs to CT10~\cite{Lai:2010vv} which does use these
data, while HERAPDF1.0 has recently updated to
HERAPDF1.5~\cite{CooperSarkar:2010wm} using an preliminary combination
of HERA data from 2003--2007 as well as the published combined
data. We will utilise the CT10 and HERAPDF1.5 PDFs for the present
study; we also consider the MSTW2008 PDFs in order to compare with
other recent calculations of high energy neutrino
cross-sections~\cite{Connolly:2011vc}, although we caution that these
have \emph{not} included the most accurate HERA low $x$ data relevant
to the present study.

PDFs are generally determined by assuming a parameterization in $x$
which is valid at a starting value of $Q^2=Q^2_0$, where the value of
$Q^2_0$ is chosen to be sufficiently large that perturbative QCD
calculations can be applied. The form of the parameterization is
usually $A x^B(1-x)^C P(x)$, where $P(x)$ is some smooth function of
$x$.  Such forms are assumed for the light quarks and the gluon,
whereas heavy quarks are generated dyamically from boson-gluon
fusion. The PDFs at all other scales $Q^2 > Q^2_0$ are calculated by
using the DGLAP equations to evolve the parameterized forms in
$Q^2$. The evolved PDFs are then convoluted with NLO matrix elements
to calculate scattering cross-sections for processes of interest.
These cross-section predictions are then fitted to deep inelastic data
over a broad range of the $x,Q^2$ plane (HERA data cover 5 decades in
both $x$ and $Q^2$) in order to determine the parameters of the
starting PDF parameterization.~\footnote{Note that not all parameters
  are fitted: the normalisations of the gluon and the $u$ and $d$
  valence quarks are fixed by the momentum and number sum rules
  respectively while some other parameters are fixed by model
  assumptions; the individual publications of the PDF groups should be
  consulted for details~\cite{Alekhin:2011sk}.}  In this way a small
number of parameters (10--25, depending on the PDF set) are fitted to
a large number (1000--2000) of data points.  In this approach one can
predict PDFs for $x$ values below those for which data exists because
the fitted functional form may be applied for all $x$, although
uncertainties will naturally increase outside the fitted region.
As we will illustrate, the uncertainty depends mainly on the  
theoretical prejudice underlying the parameterization at low $x$.

The HERA data form the back-bone of PDF fits and are the \emph{only}
data which extend to the low $x$ region. HERAPDFs use exclusively
these data, while MSTW and CT also use older, fixed target data,
Drell-Yan data including $W$ and $Z$ production, and Tevatron jet
data. Moreover HERAPDFs are unique in using only proton data, so they
are free of any assumptions concerning heavy target corrections and
deuterium binding corrections~\cite{Accardi:2011fa}.
 
PDFs are presented accounting for the correlated systematic errors of
the data as well as the statistical and uncorrelated sources, however,
each PDF group has its own approach to the estimation of confidence
limits on the uncertainties.  A general discussion of approaches to
estimating PDF uncertainties is given in
Refs.~\cite{Chekanov:2002pv,CooperSarkar:2002yx} and the approaches
used in the PDF sets considered here are reviewed in the PDF4LHC
document~\cite{Alekhin:2011sk}. PDF uncertainties can also arise from
input assumptions made in the PDF fitting. These include the form of
the input PDF parameterization at the starting scale for evolution,
$Q^2_0$, the value of $Q^2_0$ itself, the kinematic cuts made on the
data entering the fit, the value of $\alpha_\mathrm{s} (M_Z)$, the
values of the heavy quark masses --- or even the scheme used to
account for heavy quark production within the DGLAP fits. The PDFs
considered here all account for the heavy quark production using
general mass variable flavour number schemes, although the specific
schemes differ: MSTW2008 and HERAPDF1.5 use the Thorne-Roberts scheme
\cite{Thorne:1997ga,Thorne:2006qt}, whereas CT10/CTEQ6.6 use the ACOT
scheme \cite{Kramer:2000hn}.

The experimental uncertainties on the PDFs are presented as
eigenvector error sets.  These eigenvectors represent linear
combinations of the PDF parameters which are uncorrelated with each
other since they are obtained by diagonalisation of the error matrix
of the fitted parameters. Thus uncertainties on the PDFs, and
quantities derived from them, can be calculated simply from adding in
quadrature the difference between the independent eigenvector sets and
the central PDF set.\footnote{We have used Eq.(19) of
  Ref.~\cite{Campbell:2006wx}, which allows for asymmetry of the
  positive and negative excursions along the same eigenvector} However
the confidence limits represented by these PDF uncertainties are not
always set using the conventional tolerances of $\Delta\chi^2=1$ for a
68\% c.l., and $\Delta\chi^2=2.7$ for a 90\% c.l. Instead both CT(EQ)
and MSTW \emph{increase} the tolerances to account for the marginal
inconsistencies of some of the input data sets, and for possible
parameterization bias.  The exact value of the tolerance is different
for each eigenvector (see the individual PDF publications for
details). The average tolerance for $90\%$ c.l. is $\Delta\chi^2\sim
5$ for MSTW2008 and $\Delta\chi^2\sim 10$ for CTEQ6.6 and CT10.  Note
that MSTW supply both $68\%$ and $90\%$ c.l. sets, whereas
CT10/CTEQ6.6 provide only $90\%$ c.l. sets; since we wish to consider
$68\%$ c.l. uncertainties, we simply scale the CT10 error sets by
$(1.64)^{-1}$.

The HERAPDFs use $\Delta\chi^2=1$ to set the size of their $68\%$
c.l. experimental errors. The combination of the HERA data results in
a consistent data set with small, well understood systematic
errors~\cite{Aaron:2009wt}.  The HERAPDF1.0 used NC and CC $e^+$ and
$e^-$ cross-section data from the first phase of HERA running
1992--2000 (HERA-I), while the HERAPDF1.5 update uses the preliminary
combination of both HERA-I and HERA-II (2003--2007) running.  Both of
these HERAPDF sets provide not only PDF eigenvector error sets for
experimental errors but also PDF variation sets which account for
variation of input assumptions: namely variation of the charm and
beauty masses, variation of the strangeness fraction in the sea,
variation of the $Q^2$ cuts applied to the data, variation of the
parametrization at $Q^2_0$ and of the value of $Q^2_0$ itself. Not all
of these sources of uncertainty are specifically considered by the
other PDFs.

A potentially important uncertainty is the variation of the minimum
$Q^2$ cut for data entering the fit. This is set at $Q^2 >
3.5$~GeV$^2$ for HERAPDF1.5 and at $Q^2 > 2.5$~GeV$^2$ for CT10 and
MSTW2008. Although this cut should be high enough to avoid the
non-perturbative region, it may include low $x$ data for which BFKL,
or other beyond-DGLAP effects, are already important. Hence we vary
this cut to investigate the possible bias thus introduced.

More importantly, HERAPDFs consider a variant of the input
parameterization of the gluon at the starting scale which allows the
gluon to become negative at low $x$, $Q^2$.  At NLO the gluon PDF does
not have to be positive, although one might consider that it going
negative signals a breakdown of the DGLAP formalism. This form of the
gluon is in fact standard in the MSTW2008 parameterization; by
contrast, the CT(EQ) analyses do not allow such negative gluon
variants. However measurable quantities such as the longitudinal
structure function $F_L$, which is closely related to the gluon at
small $x$, \emph{must} be positive. Therefore we check whether this is
the case when a negative gluon PDF is encountered.

A further uncertainty concerns the choice of the value of
$\alpha_\mathrm{s} (M_Z)$ because the shape of the low $x$ gluon and
the value of $\alpha_\mathrm{s} (M_Z)$ are correlated through the
DGLAP equations.  All the PDF sets considered have been determined
with a fixed value of $\alpha_\mathrm{s} (M_Z)$ --- 0.120 for
MSTW2008, 0.118 for CT/CTEQ and 0.1176 for HERAPDF1.5.  However PDF
sets with a range of fixed values of $\alpha_\mathrm{s} (M^2_Z)$ are
also supplied for all these PDFs.  The uncertainty on the PDFs due to
$\alpha_\mathrm{s}(M_Z)$ can be obtained by varying the value of
$\alpha_\mathrm{s}(M_Z)$ by $\pm 0.0012$ at
$68\%$c.l. \cite{Botje:2011sn}.

The PDFs discussed above are calculated at NLO in the DGLAP formalism
and when they are used to make predictions of the neutrino DIS
cross-sections the calculations should naturally be performed at a
consistent order as we have done. We have also checked the impact of
the use of NNLO PDFs and evolution for HERAPDF1.5 and found it to be
negligible.

\section{Technical details}

The calculation of the CC and NC cross-sections in NLO has been
performed using \texttt{DISPred}~\cite{Ferrando:2010dx}.
\texttt{QCDNUM}~\cite{Botje:1999dj} as included in LHAPDF is used for
the evaluation of the structure functions $F_2$, $F_L$ and $xF_3$.
The numerical integration of the differential cross-section is
performed using the \texttt{VEGAS} Monte Carlo integration routine
from the \texttt{GSL} library~\cite{Galassi:2009aa}, reducing the
integration error to less than 1\%.

The most up-to-date PDFs in the LHAPDF~\cite{LHAPDF} format are given
in form of grids with a limited kinematic range. For example, the grid
extends down to $10^{-8}$ in $x$ for the CT10 and HERAPDF1.5 sets and
down to only $10^{-6}$ for the MSTW2008 set. Below this range the PDFs
`freeze', i.e. they take the value at the lower grid
boundary. However, for very high energies the total cross-section has
significant contributions from lower $x$ (see also
Fig.~\ref{fig:kinreg}) such that the cross-section calculated using
the LHAPDF sets will be too \emph{small}. Therefore, we have used
implementations for the PDFs provided by the CTEQ~\cite{CTEQ} and
MSTW~\cite{MSTW} groups which allow an extrapolation beyond the grid.
Of course, such an extrapolation using e.g. polynomials is not
necessarily physical but in the absence of a grid that extends to low
enough values of $x$ or the parametric form of the PDFs, this is the
only viable alternative. We have checked the agreement between the
cross-sections calculated with `freezing' and with the extrapolations
and the difference is more than 50\% for MSTW2008 but only 3\% for
CT10 at $E_{\nu} = 5 \times 10^{11}$~GeV.

\section{Results}

\begin{figure}[tbp]
\includegraphics[scale=0.95]{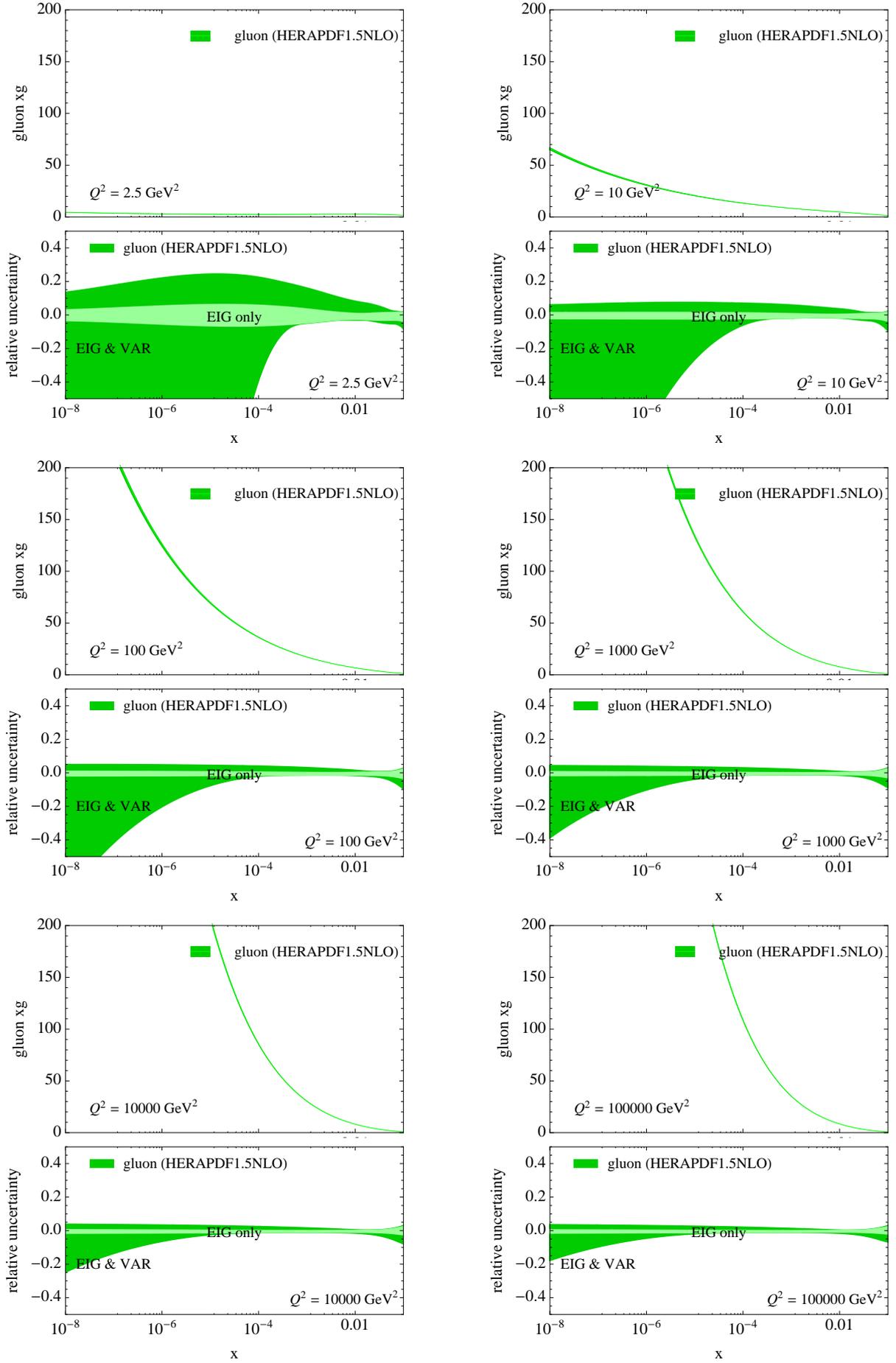}
\caption{The HERAPDF1.5 gluon PDFs and their fractional uncertainties
  --- from the experimental errors only (EIG), and from both
  experimental errors as well as model and parameter variations
  (EIG \& VAR) --- at various values of $Q^2$.}
\label{fig:glu}
\end{figure}

Figure~\ref{fig:glu} shows the predicted sea and gluon distributions
from the HERAPDF1.5 fit and their fractional uncertainties, at various
$Q^2$ values. This shows the importance of the low $x$ contribution
and illustrates that the PDF uncertainties are largest at low $Q^2$
and at low $x$. PDF uncertainties are also large at very high $x$ but
this kinematic region is not important for scattering at high neutrino
energies.

\begin{figure}[tbp]
\includegraphics[width=\textwidth]{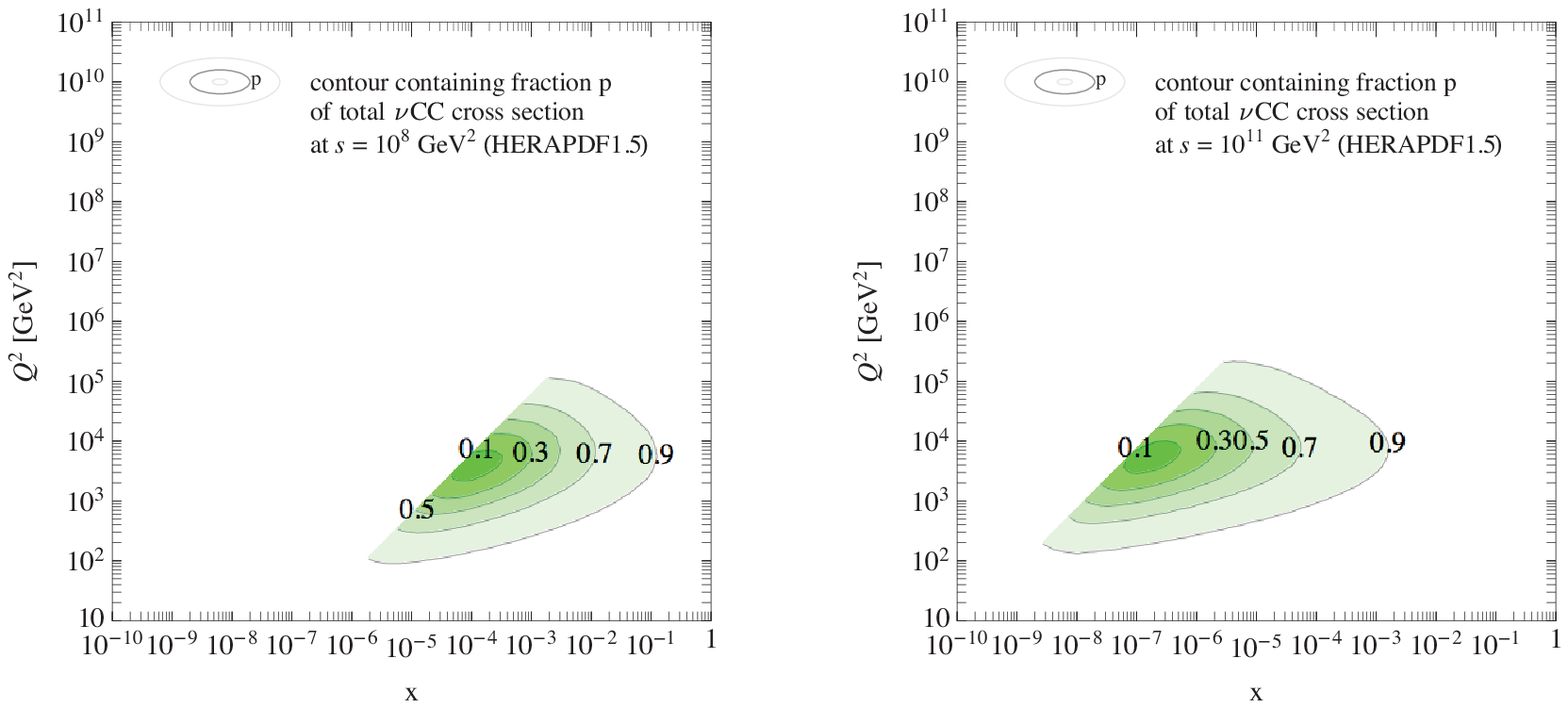}
\caption{Kinematic regions of the $x, Q^2$ plane and their
  contribution to the total neutrino cross-section using HERAPDF1.5
  for two different values of $s$. The labels show the relative
  contribution to the total cross-section contained within each
  contour.}
\label{fig:kinreg}
\end{figure}

This is seen from Figure~\ref{fig:kinreg} which indicates the
kinematic regions in the $x,Q^2$ plane which contribute to the
neutrino cross-sections for two representative values of the neutrino
energy: $s= 10^8$~GeV$^2$ ($\Rightarrow E_\nu=5.3\times 10^7$~GeV) in
the left panel and $s= 10^{10}$~GeV$^2$ ($\Rightarrow E_\nu=5.3 \times
10^9$~GeV) in the right panel. (We do not show the antineutrino
cross-sections separately because these are very close to the neutrino
cross-sections at high energy.)  One can see that the dominant
contributions come from $500 \lesssim Q^2 \lesssim 50000~$GeV$^2$ and
$10^{-6} \lesssim x \lesssim 10^{-2}$ for the lower neutrino energy,
and $10^{-8} \lesssim x \lesssim 10^{-4}$, for the higher neutrino
energy. Thus although PDF uncertainties are large at low $x$ and low
$Q^2$, we see that the dominant contributions to the cross-section do
not come from the kinematic regions of greatest uncertainty for the
PDFs.

\begin{figure}[tbp]
\includegraphics[width=\textwidth]{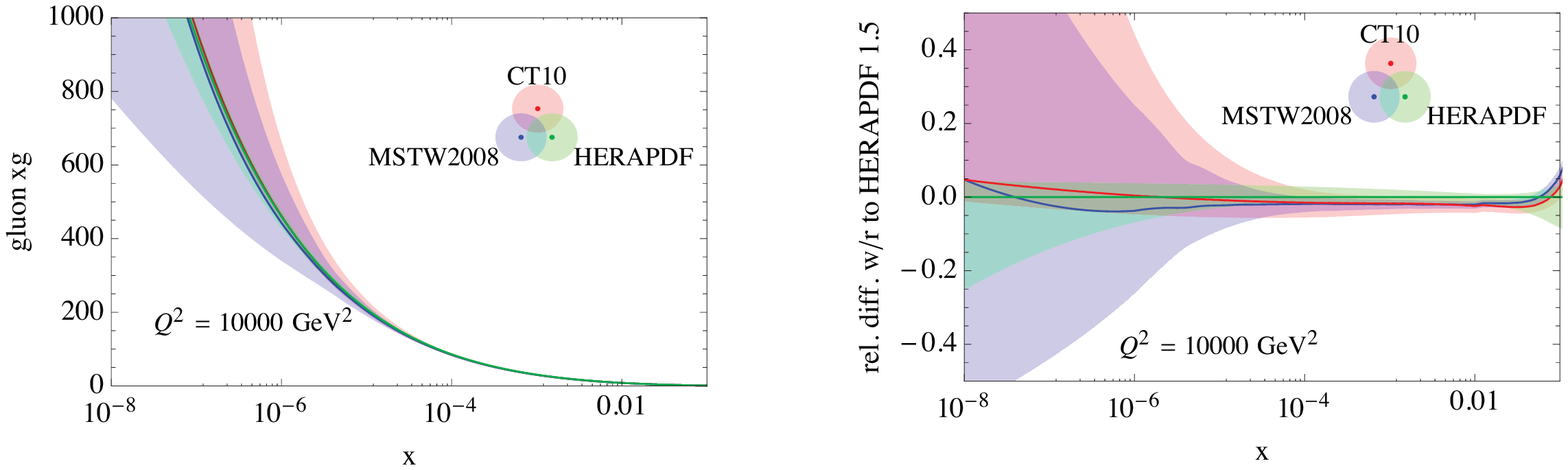}
\caption{{\bf Left panel:} Gluon structure function at $Q^2 =
  10^4\,\text{GeV}^2$ for the three PDF sets used. {\bf Right panel.}
  The relative deviations and uncertainties (at 68\% c.l.) with
  respect to the central value of HERAPDF1.5. The uncertainty bands
  are shown \emph{with} member 9 for HERAPDF1.5 and member 52 for
  CT10.}
\label{fig:glucomp}
\end{figure}

\begin{figure}[tbp]
\includegraphics[width=\textwidth]{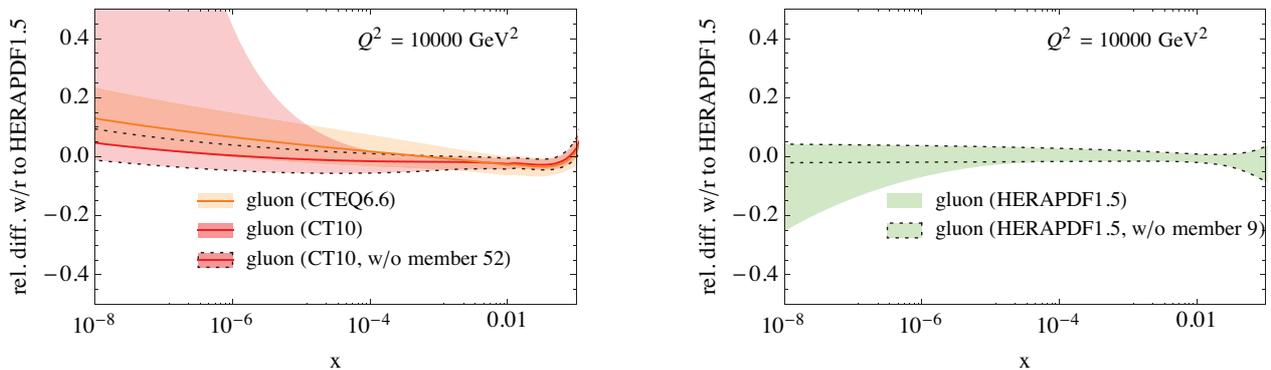}
\caption{Uncertainties (at 68\% c.l.) with respect to the central
  value of HERAPDF1.5 at $Q^2 = 10^4\,\text{GeV}^2$. {\bf Left panel:}
  CT10 with and without member 52 (CTEQ6.6 shown for comparison). {\bf
    Right panel:} HERAPDF1.5 with and without member 9. }
\label{fig:glucompnegative}
\end{figure}

Figure~\ref{fig:glucomp} compares the gluon PDF and its uncertainty at
$Q^2=10000$~GeV$^2$ for the three PDFs which we consider. This value
of $Q^2$ is in the middle of the range which contributes significantly
to the neutrino cross-sections. We see that the central values of the
gluon PDFs are all very similar, whereas the uncertainty estimates
differ.  The CT10 and HERAPDF1.5 uncertainties are actually very
similar if we leave out member 52 from the CT10 error set. This error 
set was introduced into the CT10 analysis to allow for a  larger uncertainty 
at low $x$~\cite{P.Nadolskyprivatecomm}. Previous CTEQ analyses such
as CTEQ6.6 \cite{Tung:2006tb} do not have such an extreme error set
--- see left panel of Fig.~\ref{fig:glucompnegative}. The problem with
such an \emph{ad hoc} introduction of a steeply increasing gluon PDF
is that at low $x$ it leads to a very strong rise of the neutrino
cross-section which seems unphysical (see later discussion).

\begin{figure}[tbp]
\includegraphics[width=0.49\textwidth]{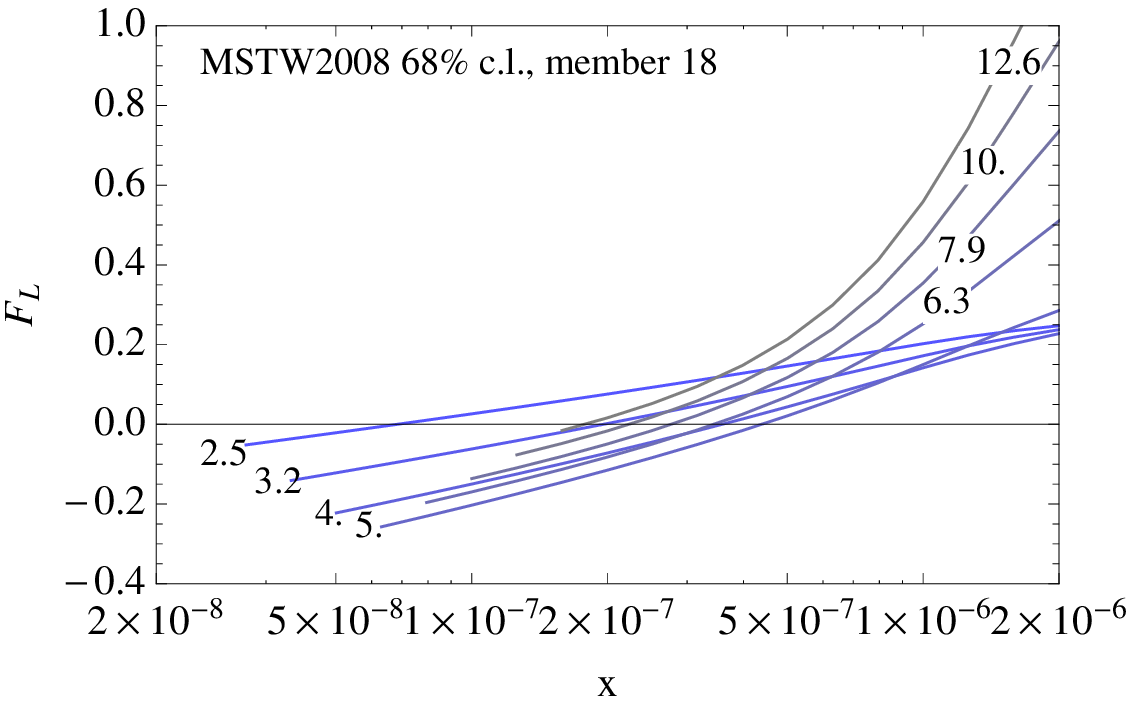}
\includegraphics[width=0.49\textwidth]{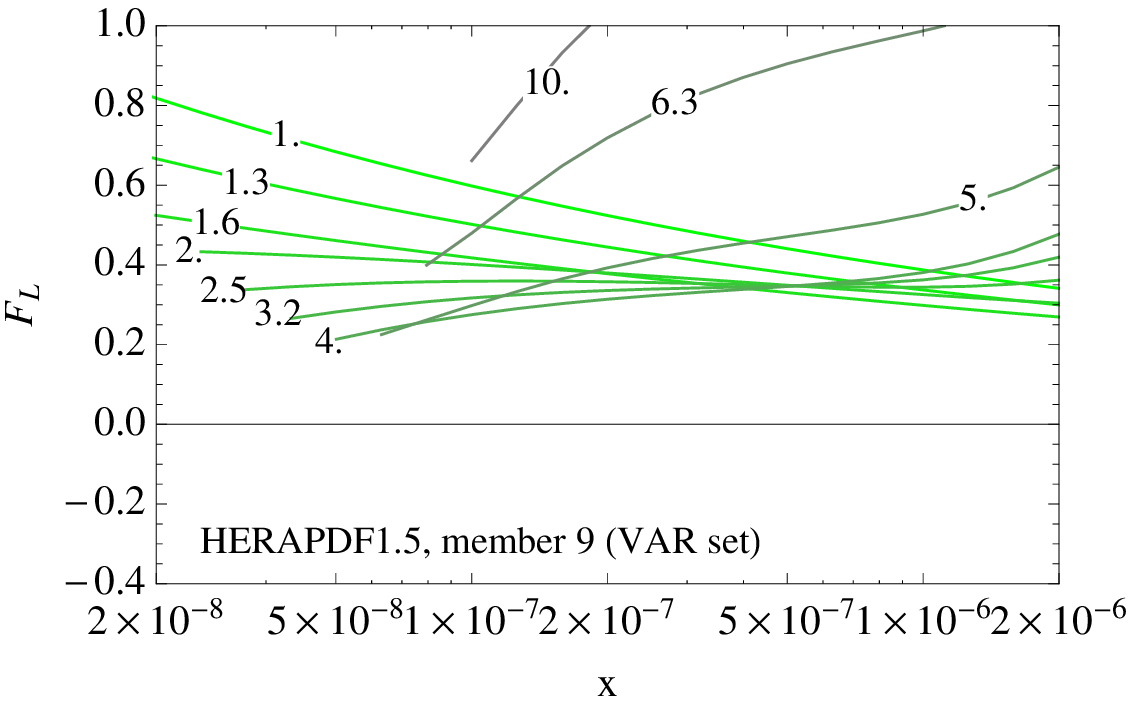}
\caption{{\bf Left panel:} The longitudinal structure function $F_L$
  calculated from member 18 of MSTW2008nlo68cl for $s =
  10^8\,\mathrm{GeV}^2$ and different values of $Q^2$ (labelled in
  units of $\mathrm{GeV}^2$).  {\bf Right panel:} The same for member
  9 (VAR set) of HERAPDF1.5.}
\label{fig:FLneg}
\end{figure}

The larger error band of MSTW2008 is partly due to the fact that it
does not include the most up to date HERA data, which have
significantly reduced errors at low $x$. However the more striking
difference between MSTW2008 and both HERAPDF1.5 and CT10 is the
downward divergence of its error band which is due to the gluon
becoming negative at low $x,\,Q^2$. This is best understood by
reference to the right panel of Fig.\ref{fig:glucompnegative} which
shows HERAPDF1.5 both with and without member 9 (the variant which
allows the gluon to become negative at low $x,\,Q^2$). However
Fig.~\ref{fig:FLneg} shows that while the longitudinal structure
function $F_L$ is always positive for HERAPDF1.5, it becomes
\emph{negative}, hence unphysical, for some of the MSTW2008 PDF error
sets.


\begin{figure}[tbp]
\includegraphics[width=\textwidth]{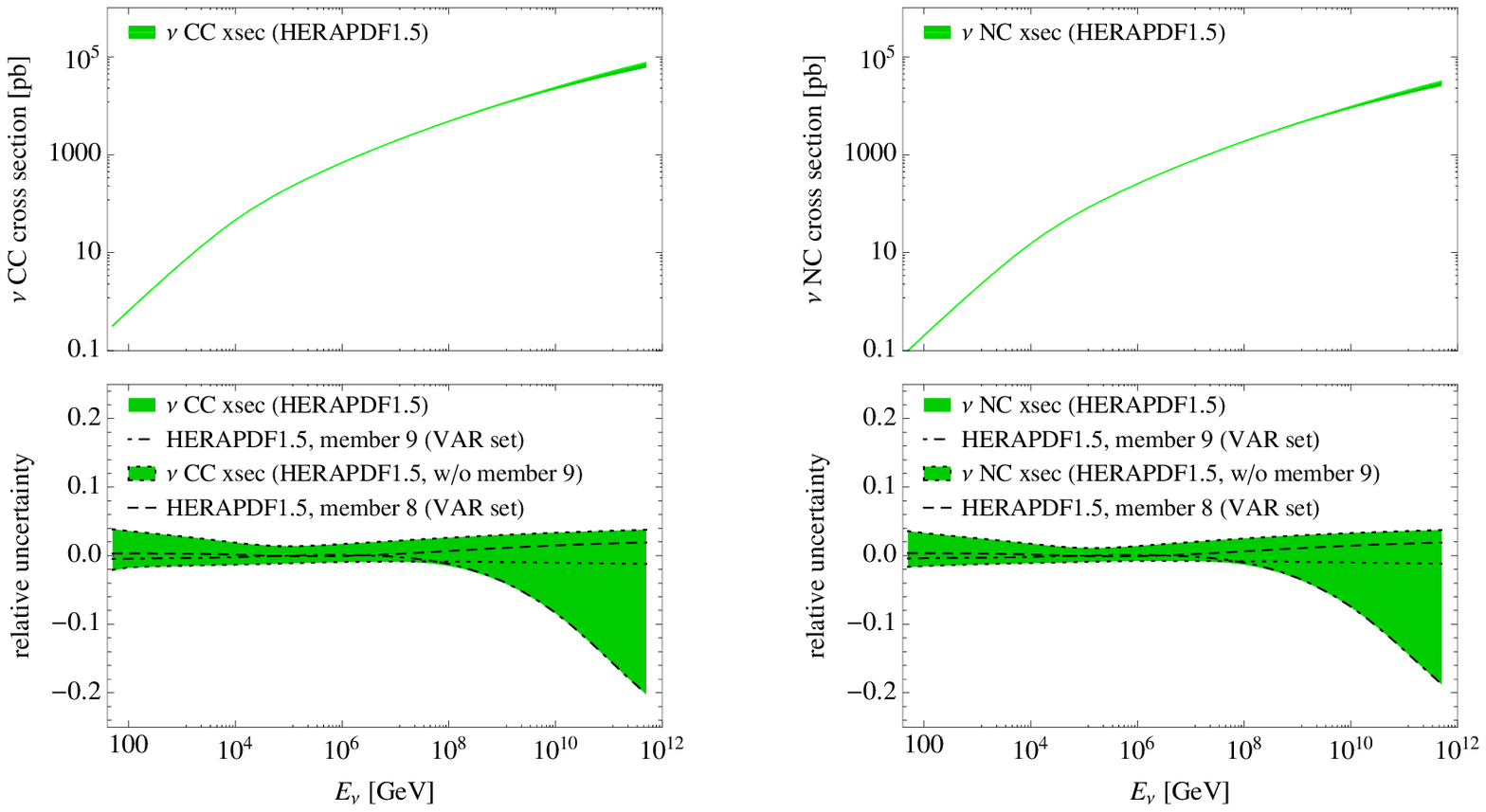}
\caption{Neutrino DIS cross-section for CC and NC scattering as
  predicted by the HERAPDF1.5. Total relative PDF uncertainties
  including experimental errors as well as model and parameter
  variations (EIG \& VAR) are shown beneath each plot, both with and
  without member 9. The marginal effect of increasing the $Q^2$ cut
  from 2.5 to 5~GeV$^2$ is also shown (member 8).}
\label{fig:xsecHERA}
\end{figure}

\begin{figure}[tbp]
\includegraphics[width=\textwidth]{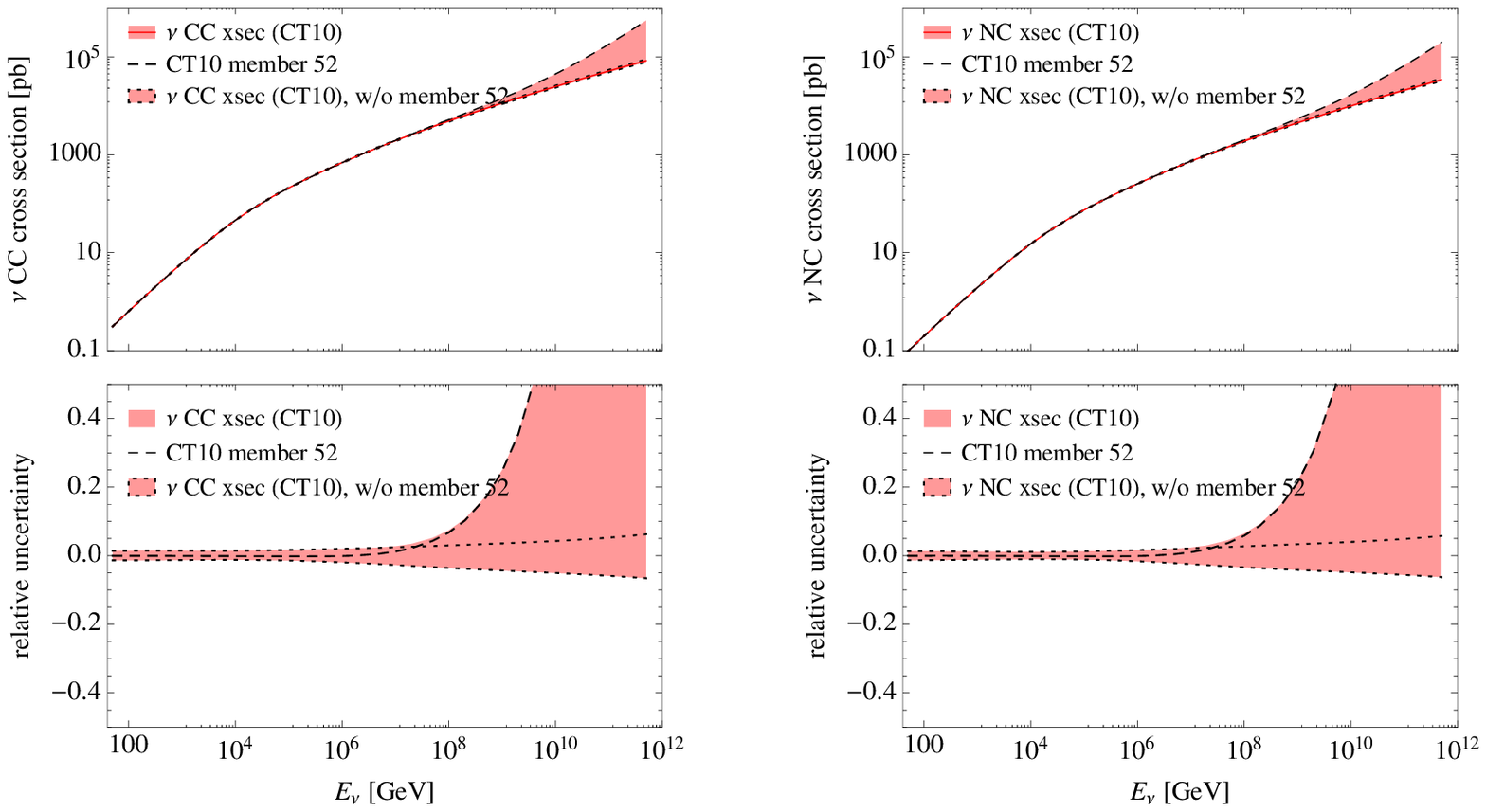}
\caption{Neutrino DIS cross-section for CC and NC scattering as
  predicted by the CT10 PDF. Relative PDF uncertainties from
  experimental errors only (EIG), and also including parameter
  uncertainties (EIG \& VAR) are shown beneath each plot, both with
  and without member 52. }
\label{fig:xsecCT10}
\end{figure}

The total neutrino cross-sections are now obtained by integrating the
predicted double differential cross-section
$\mathrm{d}^2\sigma/\mathrm{d}x\mathrm{d}y$ with no cuts on either
kinematic variable. Fig.~\ref{fig:xsecHERA} and~\ref{fig:xsecCT10}
show the NC and CC neutrino cross-sections as a function of $E_\nu$ as
evaluated from the HERAPDF1.5 and CT10 PDFs respectively. The PDF
uncertainties of these predictions are shown relative to the central
values underneath each plot.

It is clear already that the variation of input assumptions considered
by the HERAPDF1.5 --- apart from the negative gluon variation --- do
not increase the error band very significantly. Thus the other
parameterization variations and the input values of the charm and
beauty mass and the fraction of strangeness in the sea are not
important for the predictions of neutrino cross-sections. However it
is interesting to consider the variation due to the choice of the
minimum $Q^2$ for data entering the fit. Decreasing this cut to
$Q^2>2.5$~GeV$^2$ has negligible effect but the effect of increasing
the cut to $Q^2> 5$~GeV$^2$ is just about visible in
Fig.~\ref{fig:xsecHERA}. This larger $Q^2$ cut also cuts out data at
$x < 10^{-4}$ --- the kinematic region where there are already some
hints of beyond-DGLAP behaviour such as BFKL or non-linear
effects. This results in a steeper low $x$ gluon and we can see this
in the marginally enhanced neutrino cross-section.
 
Another potentially important effect comes from the variation of
$\alpha_\mathrm{s} (M_Z)$ which is correlated to the gluon PDF such
that lower values of $\alpha_\mathrm{s} (M_Z)$ result in a steeper low
$x$ gluon. We evaluate this by considering a 68\% c.l. variation of
$\Delta\alpha_\mathrm{s} (M_Z) \pm 0.0012$ from its central value.
The slight enhancement in the neutrino DIS cross-section is so small
that it is not noticeable on the scale of Figs.~\ref{fig:xsecHERA} and
\ref{fig:xsecCT10} hence we have not attempted to show it.

\begin{figure}[p]
\includegraphics[width=\textwidth]{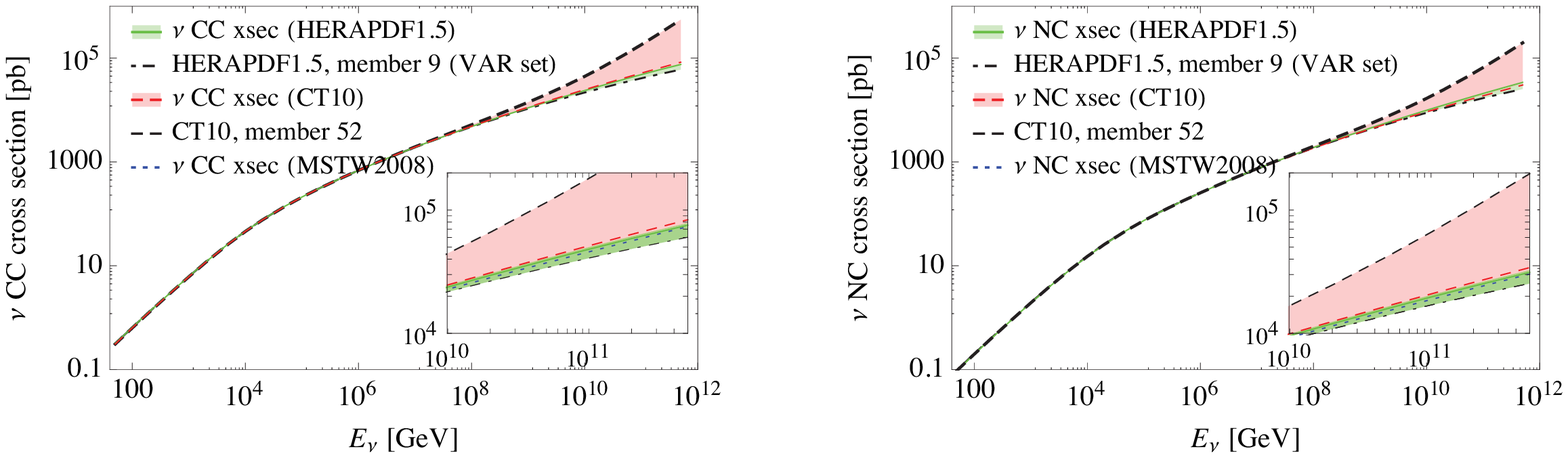}
\includegraphics[width=\textwidth]{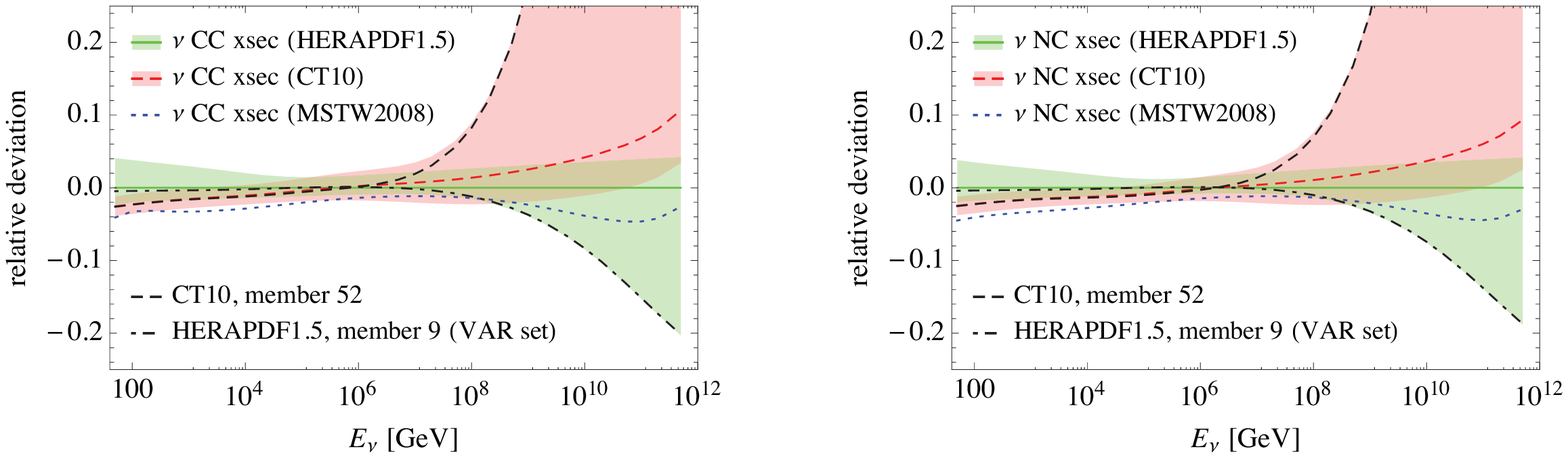}
\caption{Comparison of the total cross-section (top panels) and
  uncertainties (bottom panels) for CC and NC scattering as predicted
  by the HERAPDF1.5, CT10 and MSTW2008 (central member only) PDF sets.
  The cross-sections and deviations for member 9 of HERAPDF1.5 and
  member 52 of CT10 are indicated by the dashed and dot-dashed lines,
  respectively.}
\label{fig:xsecAllWith}
\end{figure}

\begin{figure}[p]
\includegraphics[width=\textwidth]{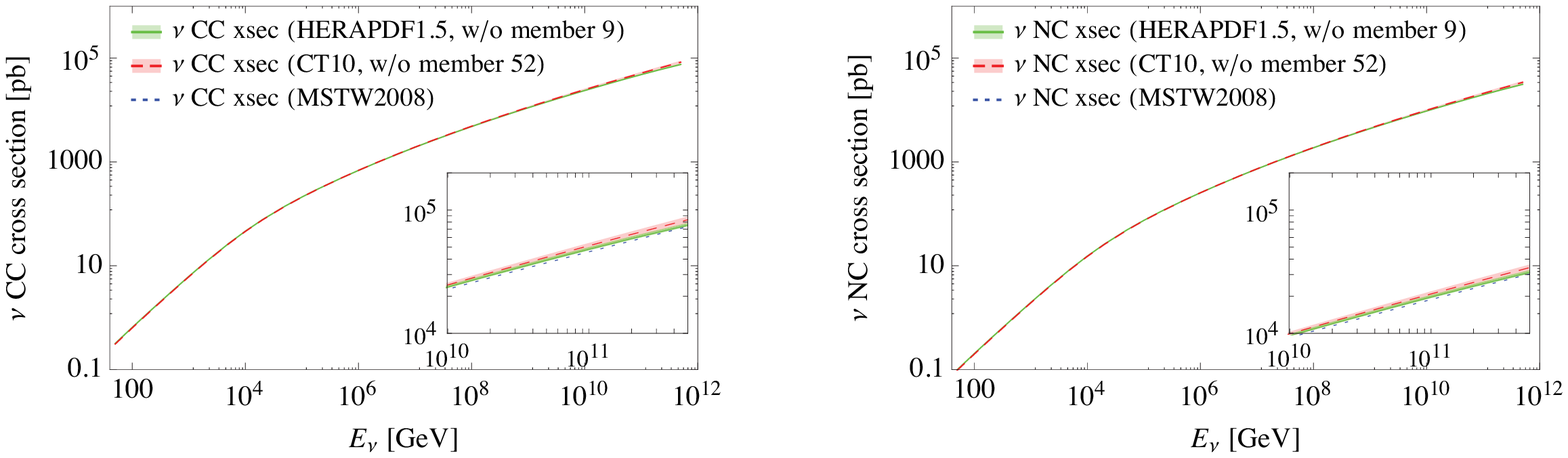}
\includegraphics[width=\textwidth]{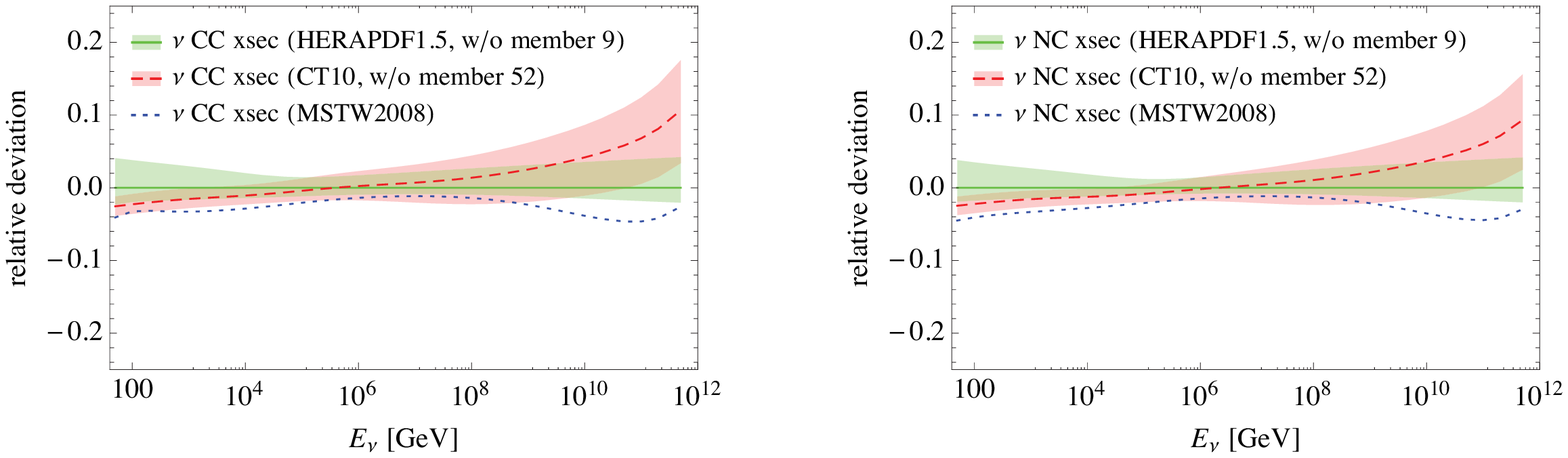}
\caption{Same as Fig.~\ref{fig:xsecAllWith}, but excluding member 9 of
  the HERAPDF1.5 set and member 52 of the CT10 set.}
\label{fig:xsecAllWithout}
\end{figure}

In Figs~\ref{fig:xsecAllWith} (top panels) we compare the NC and CC
cross-sections, along with their total uncertainties (including that
coming from the variation of $\alpha_\mathrm{s}(M_Z)$), as predicted
by HERAPDF1.5 and CT10. The MSTW2008 central prediction is also
included for comparison. In Fig~\ref{fig:xsecAllWith} (bottom panels)
we emphasize the small differences in the central values of the PDFs
and their relative uncertainties. In order to highlight the effect of
the extreme members of HERAPDF1.5 and CT10 in
Figs~\ref{fig:xsecAllWithout}, we show these plots without member 9 of
the HERAPDF15 variations (which allows for the gluon to become
negative at low $x$ and $Q^2$) and without member 52 for CT10 (the
cross-section for which rises $\propto E_\nu^{0.7}$ whereas for the
central member it rises $\propto E_\nu^{0.3}$). However \textit{any}
power-law rise in the cross-section will eventually violate the
Froissart bound, which requires the rise to be no faster than $\log^2
s$~\cite{Fiore:2005wf}. This should result in a reduction of the
cross-section at high energies, by a factor of $\sim 2$ at $E_\nu =
10^{12}$~GeV~\cite{Block:2010ud} and perhaps even more
\cite{Illarionov:2011wc}.


\begin{figure}[tbp]
\includegraphics[width=\textwidth]{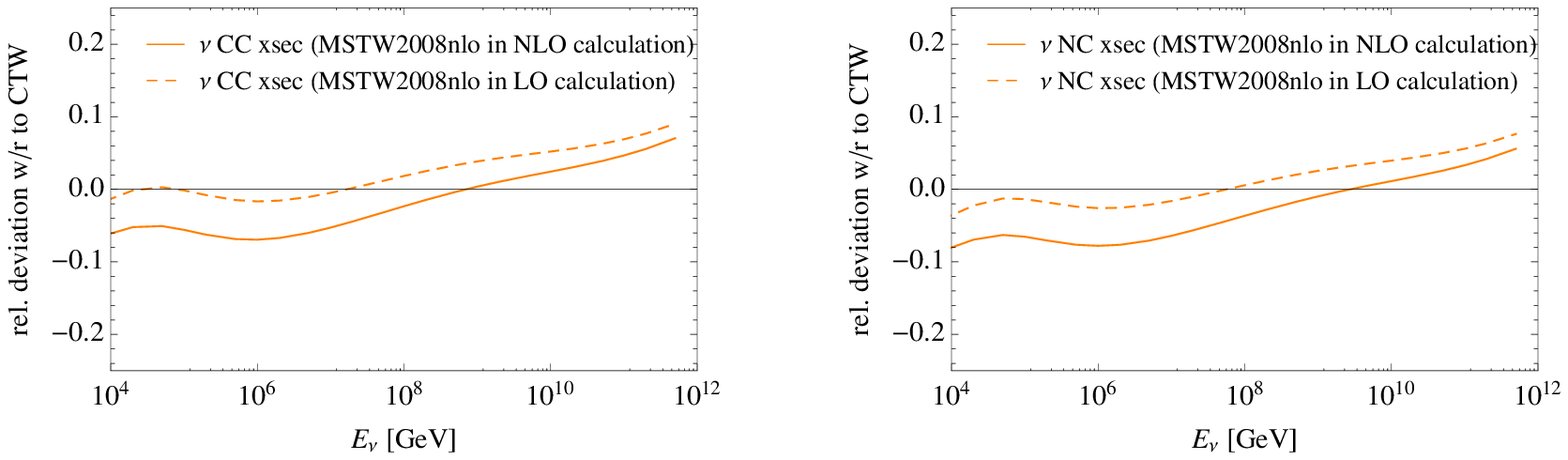}
\caption{Our results for the neutrino DIS CC and NC cross-section
  using the MSTW2008 central member, relative to the results of
  CTW~\cite{Connolly:2011vc}. The calculation is done consistenly in
  NLO as well as, for illustration, using a LO code.}
\label{fig:CTW}
\end{figure}

\begin{figure}[tbp]
\includegraphics[width=\textwidth]{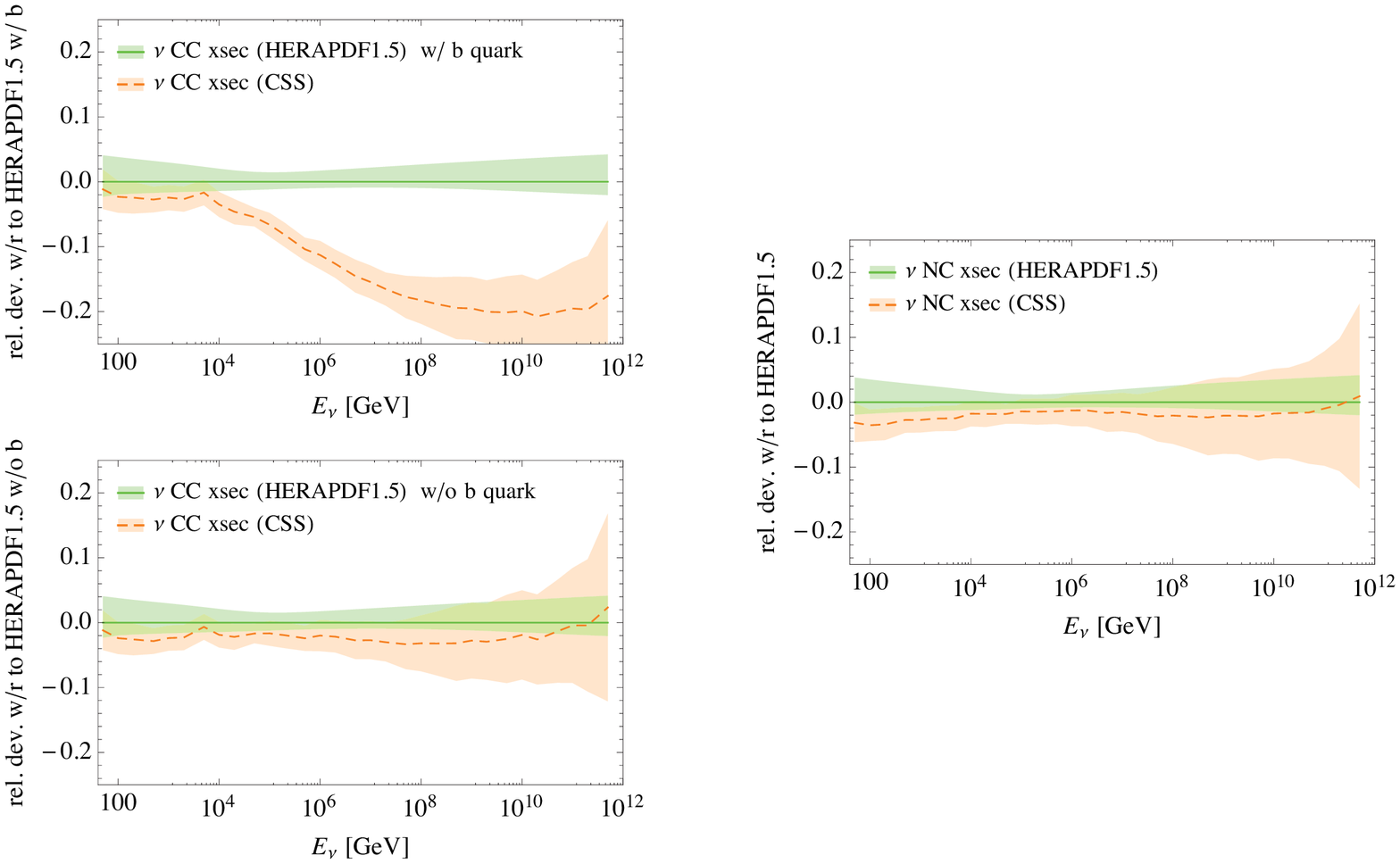}
\caption{The relative deviation of the cross-section calculated by
  CSS~\cite{CooperSarkar:2007cv} from our result for the HERAPDF1.5
  central member. For the CC cross-section (left) we compare to
  HERAPDF1.5 with (upper panel) and without (lower panel) the
  $b$-quark contribution. For NC scattering (right) the $b$-quark was
  included already by CSS~\cite{CooperSarkar:2007cv} and the agreement
  is excellent.}
\label{fig:CSS}
\end{figure}

There are small differences from the previous work of
CSS~\cite{CooperSarkar:2007cv} and CTW~\cite{Connolly:2011vc}.  In
Fig.~\ref{fig:CTW} we compare our calculation using MSTW2008 PDFs to
that of CTW.  We find that we can reproduce their results well only if
we use the MSTW2008 NLO PDFs together with an \emph{leading-order}
treatment of the coefficient functions, rather than with a consistent
NLO approach.
In Fig~\ref{fig:CSS} we compare our calculation using HERAPDF1.5 to
that of CSS \cite{CooperSarkar:2007cv}. The predictions for both the
central values and the uncertainties of the neutrino NC cross-section
are quite close. This is also true for the CC cross-section at low
$E_\nu$, however above $10^4$~GeV the difference increases since in
the present work we have included the contribution of the $b$ quark
which was missed out in our earlier work. Accordingly the bounds
derived on the cosmogenic neutrino flux using the CSS cross-sections
\cite{Abbasi:2010ak,Abraham:2007rj} are conservative, being $\sim
20\%$ too high.

\begin{figure}[tbp]
\includegraphics[width=\textwidth]{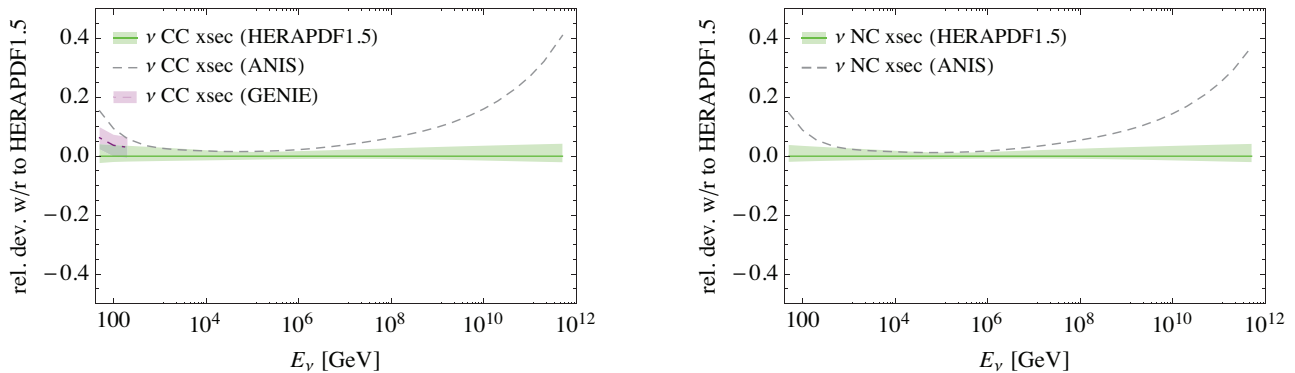}
\caption{The relative deviation of the ANIS~\cite{Gazizov:2004va} and
  GENIE~\cite{Andreopoulos:2009rq} cross-sections from the HERAPDF1.5
  central member.}
\label{fig:ANIS}
\end{figure}

In Fig.~\ref{fig:ANIS} we compare our results for HERAPDF1.5 to the
cross-sections used in the neutrino event generator
\texttt{ANIS}~\cite{Gazizov:2004va} which is based on CTEQ5D. Note
that at energies below a TeV (which is the most important energy range
for neutrino telescopes like IceCube~\cite{Abbasi:2010ie}) there is a
$\sim 10\,\%$ discrepancy. We also compare the CC cross-section for
HERAPDF1.5 to its value in the GENIE low energy neutrino event
generator~\cite{Andreopoulos:2009rq} at around 100 GeV, finding the
match to be consistent within errors.

\section{Conclusions}

We find that the predictions of high energy neutrino DIS
cross-sections from the central values of HERAPDF1.5, CT10 and
MSTW2008 PDFs are very similar. However the predictions for the
uncertainties (deriving from the uncertainties on the input PDFs)
differ quite strongly. In fact PDF uncertainties derive from the input
assumptions as well as from the input experimental data.  If we
exclude error sets which either lead to too steep a rise in the
cross-section, or allow the low $x$ gluon to be negative at low $Q^2$,
then we find that the uncertainty estimates of HERAPDF1.5 and CT10 ---
both of which use the most up-to-date, accurate HERA data --- are
remarkably consistent.

\begin{figure}[tbp]
\includegraphics[width=\textwidth]{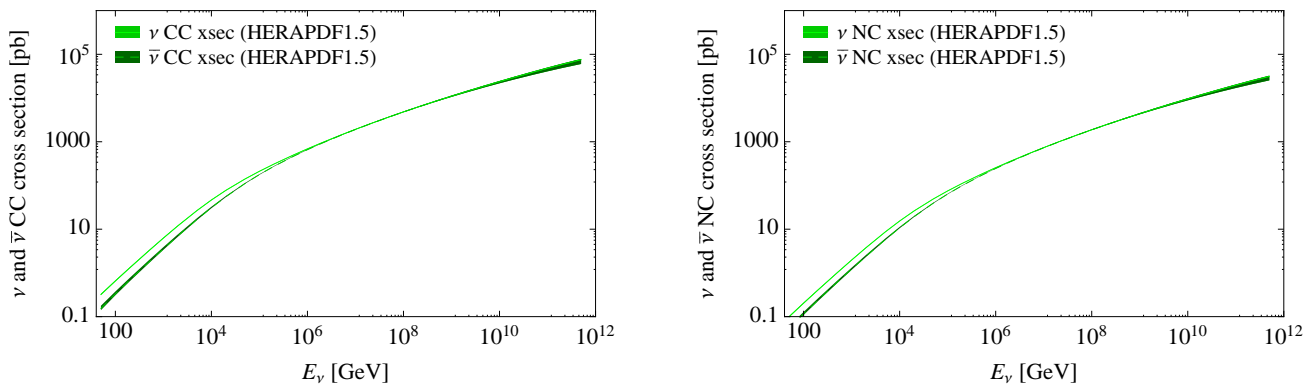}
\caption{Neutrino and anti-neutrino cross-sections on isoscalar
  targets for CC and NC scattering according to HERAPDF1.5.}
\label{fig:nuANDnubarHERA}
\end{figure}

Our results for the high energy neutrino and antineutrino CC and NC
DIS cross-sections and their uncertainties using HERAPDF1.5 at NLO are
shown in Fig.~\ref{fig:nuANDnubarHERA}. The general trend of the
uncertainties can be understood by noting that as one moves to higher
neutrino energy one also moves to lower $x$ where the PDF
uncertainties are increasing. The PDF uncertainties are smallest at
$10^{-2} \lesssim x \lesssim 10^{-1}$, corresponding to $s \sim
10^5$~GeV$^2$. Moving to smaller neutrino energies brings us into the
high $x$ region where PDF uncertainties increase again. This effect is
greater for the HERAPDF1.5 because the HERA data have less statistics
at high $x$ than the fixed target data which are included in CT10;
however these data have further uncertainties that are not fully
accounted for in CT10, e.g. heavy target corrections, deuterium
corrections and assumptions regarding higher twist effects. When the
high $x$ region becomes important the neutrino and antineutrino
cross-sections are different because the valence contribution to
$xF_3$ is now significant.  This is seen in
Fig.~\ref{fig:nuANDnubarHERA}, as is the onset of the linear
dependence of the cross-sections for $s < M_W^2$. Note that our
predictions are made for $Q^2 > 1$~GeV$^2$ since perturbative QCD
cannot sensibly be used at lower values. Moreover for $s$ below
$\sim100$~GeV$^2$, there can be contributions to the cross-section of
${\cal O}(10\%)$ from even lower values of $Q^2$ which are not
accounted for here; hence we do not show results for $E_\nu$ below 50
GeV where there are other contributions to the neutrino cross-section
and the use of a code such as GENIE ~\cite{Andreopoulos:2009rq} is
appropriate. For higher energies, we intend to upgrade ANIS
\cite{Gazizov:2004va} to use the HERAPDF1.5 (differential)
cross-sections. Meanwhile we have provided the total DIS
cross-sections for CC and NC scattering of neutrinos and antineutrinos
on isoscalar targets in Tables~\ref{tbl:nuHERA} and
\ref{tbl:nubarHERA} and recommend these as a benchmark for use by
experimentalists. These cross-sections as well as those for isoscalar 
targets are available from a webpage \cite{Mandywebpage}; differential 
cross sections are available upon request. Any measured
deviation from these values would signal the need for new physics
beyond the DGLAP formalism.

\section{Acknowledgements}

We are grateful to James Ferrando for providing us with an up-to-date
version of \texttt{DISPRED} and for discussions. We also thank Mike
Whalley and Voica Radescu for the speedy implementation of HERAPDF1.5
in LHAPDF. PM and SS thank their colleagues in the Auger and IceCube
collaborations for stimulating exchanges and acknowledge partial
support from the EU Research \& Training Network ``Unification in the
LHC era'' (PITN-GA-2009-237920).

\begin{table}[p]
	\centering
		\begin{tabular}{@{} r @{\hspace{0.5cm}} r @{.} l @{\hspace{0.5cm}} r@{~\%} l @{\hspace{0.5cm}} r@{~\%} l @{\hspace{0.5cm}} r@{~\%} l @{\hspace{0.5cm}} r @{.} l @{\hspace{0.5cm}} r@{~\%} l @{\hspace{0.5cm}} r@{~\%} l @{\hspace{0.5cm}} r@{~\%} r @{}}
			\hline
			\hline
      		\multirow{2}{*}{$E_{\nu} [ \text{GeV} ]$}    & \multicolumn{2}{c}{\multirow{2}{*}{$\sigma_{\text{CC}} [ \text{pb} ]$}} &\multicolumn{2}{c}{\multirow{2}{*}{up}} & \multicolumn{2}{c}{down} & \multicolumn{2}{c}{down} & \multicolumn{2}{c}{\multirow{2}{*}{$\sigma_{\text{NC}} [ \text{pb} ]$}} & \multicolumn{2}{c}{\multirow{2}{*}{up}} & \multicolumn{2}{c}{down} & \multicolumn{2}{c}{down} \\
      		& \multicolumn{2}{c}{} & \multicolumn{2}{c}{} & \multicolumn{2}{c}{(w/o mem. 9)} & \multicolumn{2}{c}{(w/ mem. 9)} & \multicolumn{2}{c}{} & \multicolumn{2}{c}{} & \multicolumn{2}{c}{(w/o mem. 9)} & \multicolumn{2}{c}{(w/ mem. 9)}  \\
			\hline
50&    0&    32&    4.1&    &    \hspace{0.5cm} -2.3&    &    \hspace{0.5cm} -2.4&    &    0&    10&    3.8&    &    \hspace{0.5cm} -1.9&    &    \hspace{0.5cm} -2.0&     \\ 
100&    0&    65&    3.8&    &    \hspace{0.5cm} -2.0&    &    \hspace{0.5cm} -2.0&    &    0&    20&    3.5&    &    \hspace{0.5cm} -1.8&    &    \hspace{0.5cm} -1.8&     \\ 
200&    1&    3&    3.5&    &    \hspace{0.5cm} -1.8&    &    \hspace{0.5cm} -1.9&    &    0&    41&    3.2&    &    \hspace{0.5cm} -1.6&    &    \hspace{0.5cm} -1.7&     \\ 
500&    3&    2&    3.2&    &    \hspace{0.5cm} -1.7&    &    \hspace{0.5cm} -1.8&    &    1&    0&    2.9&    &    \hspace{0.5cm} -1.5&    &    \hspace{0.5cm} -1.5&     \\ 
1000&    6&    2&    3.0&    &    \hspace{0.5cm} -1.6&    &    \hspace{0.5cm} -1.7&    &    2&    0&    2.7&    &    \hspace{0.5cm} -1.4&    &    \hspace{0.5cm} -1.5&     \\ 
2000&    12&    &    2.7&    &    \hspace{0.5cm} -1.6&    &    \hspace{0.5cm} -1.6&    &    3&    8&    2.4&    &    \hspace{0.5cm} -1.3&    &    \hspace{0.5cm} -1.4&     \\ 
5000&    27&    &    2.3&    &    \hspace{0.5cm} -1.5&    &    \hspace{0.5cm} -1.5&    &    8&    6&    2.1&    &    \hspace{0.5cm} -1.3&    &    \hspace{0.5cm} -1.3&     \\ 
10000&    47&    &    2.0&    &    \hspace{0.5cm} -1.4&    &    \hspace{0.5cm} -1.4&    &    15&    &    1.8&    &    \hspace{0.5cm} -1.2&    &    \hspace{0.5cm} -1.2&     \\ 
20000&    77&    &    1.8&    &    \hspace{0.5cm} -1.3&    &    \hspace{0.5cm} -1.4&    &    26&    &    1.6&    &    \hspace{0.5cm} -1.1&    &    \hspace{0.5cm} -1.1&     \\ 
50000&    140&    &    1.5&    &    \hspace{0.5cm} -1.2&    &    \hspace{0.5cm} -1.2&    &    49&    &    1.3&    &    \hspace{0.5cm} -1.0&    &    \hspace{0.5cm} -1.1&     \\ 
100000&    210&    &    1.4&    &    \hspace{0.5cm} -1.2&    &    \hspace{0.5cm} -1.2&    &    75&    &    1.2&    &    \hspace{0.5cm} -1.0&    &    \hspace{0.5cm} -1.0&     \\ 
200000&    310&    &    1.5&    &    \hspace{0.5cm} -1.1&    &    \hspace{0.5cm} -1.1&    &    110&    &    1.2&    &    \hspace{0.5cm} -0.9&    &    \hspace{0.5cm} -0.9&     \\ 
500000&    490&    &    1.6&    &    \hspace{0.5cm} -1.0&    &    \hspace{0.5cm} -1.0&    &    180&    &    1.3&    &    \hspace{0.5cm} -0.8&    &    \hspace{0.5cm} -0.8&     \\ 
$1\times 10^6$ &       690&    &	1.7&    &	\hspace{0.5cm} -0.9&    &	 \hspace{0.5cm} -0.9&    &    260&    &    1.4&    &    \hspace{0.5cm} -0.8&    &	\hspace{0.5cm} -0.8&     \\ 
$2\times 10^6$ &       950&    &	1.9&    &	\hspace{0.5cm} -0.9&    &	 \hspace{0.5cm} -0.9&    &    360&    &    1.6&    &    \hspace{0.5cm} -0.8&    &	\hspace{0.5cm} -0.8&     \\ 
$5 \times 10^6$ &      1400&	&	 2.0&    &	 \hspace{0.5cm} -0.9&    &    \hspace{0.5cm} -0.9&    &    540&    &	1.8&    &	\hspace{0.5cm} -0.8&    &	 \hspace{0.5cm} -0.8&     \\ 
$1 \times 10^7$ &      1900&	&	 2.2&    &	 \hspace{0.5cm} -0.9&    &    \hspace{0.5cm} -0.9&    &    730&    &	2.0&    &	\hspace{0.5cm} -0.8&    &	 \hspace{0.5cm} -0.8&     \\ 
$2 \times 10^7$ &      2600&	&	 2.3&    &	 \hspace{0.5cm} -0.9&    &    \hspace{0.5cm} -1.0&    &    980&    &	2.2&    &	\hspace{0.5cm} -0.8&    &	 \hspace{0.5cm} -0.9&     \\ 
$5 \times 10^7$ &      3700&	&	 2.5&    &	 \hspace{0.5cm} -0.9&    &    \hspace{0.5cm} -1.2&    &    1400&	&	 2.4&    &	 \hspace{0.5cm} -0.9&    &    \hspace{0.5cm} -1.1&     \\ 
$1 \times 10^8$ &      4800&	&	 2.7&    &	 \hspace{0.5cm} -0.9&    &    \hspace{0.5cm} -1.5&    &    1900&	&	 2.6&    &	 \hspace{0.5cm} -0.9&    &    \hspace{0.5cm} -1.3&     \\ 
$2 \times 10^8$ &      6200&	&	 2.8&    &	 \hspace{0.5cm} -1.0&    &    \hspace{0.5cm} -2.0&    &    2400&	&	 2.7&    &	 \hspace{0.5cm} -1.0&    &    \hspace{0.5cm} -1.8&     \\ 
$5 \times 10^8$ &      8700&	&	 3.0&    &	 \hspace{0.5cm} -1.1&    &    \hspace{0.5cm} -3.0&    &    3400&	&	 2.9&    &	 \hspace{0.5cm} -1.0&    &    \hspace{0.5cm} -2.6&     \\ 
$1 \times 10^9$ &      11000&	 &    3.1&    &    \hspace{0.5cm} -1.2&    &    \hspace{0.5cm} -3.9&    &	4400&	 &    3.0&    &    \hspace{0.5cm} -1.1&    &    \hspace{0.5cm} -3.4&     \\ 
$2 \times 10^9$ &      14000&	 &    3.3&    &    \hspace{0.5cm} -1.2&    &    \hspace{0.5cm} -5.0&    &	5600&	 &    3.2&    &    \hspace{0.5cm} -1.2&    &    \hspace{0.5cm} -4.4&     \\ 
$5 \times 10^9$ &      19000&	 &    3.4&    &    \hspace{0.5cm} -1.4&    &    \hspace{0.5cm} -6.8&    &	7600&	 &    3.4&    &    \hspace{0.5cm} -1.3&    &    \hspace{0.5cm} -6.1&     \\ 
$1 \times 10^{10}$ &	24000&    &    3.6&    &    \hspace{0.5cm} -1.5&    &	\hspace{0.5cm} -8.5&    &	 9600&    &    3.5&    &    \hspace{0.5cm} -1.4&    &	\hspace{0.5cm} -7.6&     \\ 
$2 \times 10^{10}$ &	30000&    &    3.7&    &    \hspace{0.5cm} -1.6&    &	\hspace{0.5cm} -10.3&    &    12000&	&	 3.6&    &	 \hspace{0.5cm} -1.5&    &    \hspace{0.5cm} -9.3&     \\ 
$5 \times 10^{10}$ &	39000&    &    3.8&    &    \hspace{0.5cm} -1.7&    &	\hspace{0.5cm} -13.1&    &    16000&	&	 3.8&    &	 \hspace{0.5cm} -1.7&    &    \hspace{0.5cm} -11.8&     \\ 
$1 \times 10^{11}$ &	48000&    &    4.0&    &    \hspace{0.5cm} -1.8&    &	\hspace{0.5cm} -15.2&    &    20000&	&	 3.9&    &	 \hspace{0.5cm} -1.8&    &    \hspace{0.5cm} -13.9&     \\ 
$2 \times 10^{11}$ &	59000&    &    4.1&    &    \hspace{0.5cm} -1.9&    &	\hspace{0.5cm} -17.5&    &    24000&	&	 4.0&    &	 \hspace{0.5cm} -1.9&    &    \hspace{0.5cm} -16.1&     \\ 
$5 \times 10^{11}$ &	75000&    &    4.2&    &    \hspace{0.5cm} -2.0&    &	\hspace{0.5cm} -20.3&    &    31000&	&	 4.2&    &	 \hspace{0.5cm} -2.0&    &    \hspace{0.5cm} -18.8&     \\ 
\hline
\end{tabular}
\caption{Neutrino CC and NC cross-sections on isoscalar targets,  
along with their uncertainties, in the perturbative DGLAP formalism at  
NLO, using HERAPDF1.5 (both with and without member 9).}
\label{tbl:nuHERA}
\end{table}

\begin{table}[p]
	\centering
		\begin{tabular}{@{} r @{\hspace{0.5cm}} r @{.} l @{\hspace{0.5cm}} r@{~\%} l @{\hspace{0.5cm}} r@{~\%} l @{\hspace{0.5cm}} r@{~\%} l @{\hspace{0.5cm}} r @{.} l @{\hspace{0.5cm}} r@{~\%} l @{\hspace{0.5cm}} r@{~\%} l @{\hspace{0.5cm}} r@{~\%} r @{}}
			\hline
			\hline
      		\multirow{2}{*}{$E_{\nu} [ \text{GeV} ]$}    & \multicolumn{2}{c}{\multirow{2}{*}{$\sigma_{\text{CC}} [ \text{pb} ]$}} &\multicolumn{2}{c}{\multirow{2}{*}{up}} & \multicolumn{2}{c}{down} & \multicolumn{2}{c}{down} & \multicolumn{2}{c}{\multirow{2}{*}{$\sigma_{\text{NC}} [ \text{pb} ]$}} & \multicolumn{2}{c}{\multirow{2}{*}{up}} & \multicolumn{2}{c}{down} & \multicolumn{2}{c}{down} \\
      		& \multicolumn{2}{c}{} & \multicolumn{2}{c}{} & \multicolumn{2}{c}{(w/o mem. 9)} & \multicolumn{2}{c}{(w/ mem. 9)} & \multicolumn{2}{c}{} & \multicolumn{2}{c}{} & \multicolumn{2}{c}{(w/o mem. 9)} & \multicolumn{2}{c}{(w/ mem. 9)}  \\
			\hline
50&    0&    15&    15.0&    &    \hspace{0.5cm} -9.0&    &    \hspace{0.5cm} -9.0&    &    0&    05&    12.0&    &    \hspace{0.5cm} -6.4&    &    \hspace{0.5cm} -6.4& \\ 
100&    0&    33&    13.3&    &    \hspace{0.5cm} -7.4&    &    \hspace{0.5cm} -7.4&    &    0&    12&    10.7&    &    \hspace{0.5cm} -5.7&    &    \hspace{0.5cm} -5.7& \\ 
200&    0&    69&    11.9&    &    \hspace{0.5cm} -6.5&    &    \hspace{0.5cm} -6.5&    &    0&    24&    9.6&    &    \hspace{0.5cm} -5.1&    &    \hspace{0.5cm} -5.1& \\ 
500&    1&    8&    10.5&    &    \hspace{0.5cm} -5.7&    &    \hspace{0.5cm} -5.7&    &    0&    61&    8.6&    &    \hspace{0.5cm} -4.6&    &    \hspace{0.5cm} -4.6& \\ 
1000&    3&    6&    9.4&    &    \hspace{0.5cm} -5.2&    &    \hspace{0.5cm} -5.2&    &    1&    20&    7.8&    &    \hspace{0.5cm} -4.2&    &    \hspace{0.5cm} -4.2&     \\ 
2000&    7&    &    8.3&    &    \hspace{0.5cm} -4.6&    &    \hspace{0.5cm} -4.6&    &    2&    4&    7.0&    &    \hspace{0.5cm} -3.8&    &    \hspace{0.5cm} -3.8&     \\ 
5000&    17&    &    6.5&    &    \hspace{0.5cm} -3.7&    &    \hspace{0.5cm} -3.7&    &    5&    8&    5.7&    &    \hspace{0.5cm} -3.2&    &    \hspace{0.5cm} -3.2&     \\ 
10000&    31&    &    5.1&    &    \hspace{0.5cm} -3.0&    &    \hspace{0.5cm} -3.0&    &    11&    &    4.6&    &    \hspace{0.5cm} -2.7&    &    \hspace{0.5cm} -2.7&     \\ 
20000&    55&    &    3.8&    &    \hspace{0.5cm} -2.3&    &    \hspace{0.5cm} -2.3&    &    19&    &    3.6&    &    \hspace{0.5cm} -2.1&    &    \hspace{0.5cm} -2.1&     \\ 
50000&    110&    &    2.5&    &    \hspace{0.5cm} -1.7&    &    \hspace{0.5cm} -1.7&    &    39&    &    2.4&    &    \hspace{0.5cm} -1.5&    &    \hspace{0.5cm} -1.5&     \\ 
100000&    180&    &    1.9&    &    \hspace{0.5cm} -1.4&    &    \hspace{0.5cm} -1.4&    &    64&    &    1.7&    &    \hspace{0.5cm} -1.2&    &    \hspace{0.5cm} -1.2&     \\ 
200000&    270&    &    1.7&    &    \hspace{0.5cm} -1.2&    &    \hspace{0.5cm} -1.2&    &    99&    &    1.4&    &    \hspace{0.5cm} -1.0&    &    \hspace{0.5cm} -1.0&     \\ 
500000&    460&    &    1.7&    &    \hspace{0.5cm} -1.1&    &    \hspace{0.5cm} -1.1&    &    170&    &    1.4&    &    \hspace{0.5cm} -0.9&    &    \hspace{0.5cm} -0.9&     \\ 
$1\times 10^6$ &   660&    &	1.8&	&    \hspace{0.5cm} -1.0&    &	\hspace{0.5cm} -1.0&    &	240&	&	 1.5&    &	 \hspace{0.5cm} -0.8&    &    \hspace{0.5cm} -0.8&     \\ 
$2\times 10^6$ &   920&    &	1.9&	&    \hspace{0.5cm} -1.0&    &	\hspace{0.5cm} -1.0&    &	350&	&	 1.6&    &	 \hspace{0.5cm} -0.8&    &    \hspace{0.5cm} -0.8&     \\ 
$5\times 10^6$ &   1400&	&	 2.1&	 &    \hspace{0.5cm} -0.9&    &    \hspace{0.5cm} -0.9&    &    530&    &	1.9&    &	\hspace{0.5cm} -0.8&    &	 \hspace{0.5cm} -0.8&     \\ 
$1\times 10^7$ &   1900&	&	 2.2&	 &    \hspace{0.5cm} -0.9&    &    \hspace{0.5cm} -0.9&    &    730&    &	2.0&    &    \hspace{0.5cm} -0.8&    &	\hspace{0.5cm} -0.8&     \\ 
$2\times 10^7$ &   2500&	&	 2.3&	 &    \hspace{0.5cm} -0.9&    &    \hspace{0.5cm} -1.0&    &    980&    &    2.2&    &    \hspace{0.5cm} -0.8&    &	\hspace{0.5cm} -0.9&     \\ 
$5\times 10^7$ &   3700&	&	 2.5&	 &    \hspace{0.5cm} -0.9&    &    \hspace{0.5cm} -1.2&    &    1400&	&	 2.4&    &	 \hspace{0.5cm} -0.9&    &    \hspace{0.5cm} -1.1&     \\ 
$1\times 10^8$ &   4800&	&	 2.7&	 &    \hspace{0.5cm} -1.0&    &	 \hspace{0.5cm} -1.5&    &    1900&    &	2.6&    &	\hspace{0.5cm} -0.9&    &	 \hspace{0.5cm} -1.3&     \\ 
$2\times 10^8$ &   6200&	&	 2.8&	 &    \hspace{0.5cm} -1.0&    &	 \hspace{0.5cm} -2.0&    &	 2400&    &    2.7&    &    \hspace{0.5cm} -1.0&    &    \hspace{0.5cm} -1.8&     \\ 
$5\times 10^8$ &   8700&	&	 3.0&	 &    \hspace{0.5cm} -1.1&    &	 \hspace{0.5cm} -3.0&    &	 3400&    &    2.9&    &    \hspace{0.5cm} -1.0&    &    \hspace{0.5cm} -2.6&     \\ 
$1\times 10^9$ &   11000&	 &    3.1&    &    \hspace{0.5cm} -1.2&    &    \hspace{0.5cm} -3.9&    &	4400&	 &    3.0&    &	 \hspace{0.5cm} -1.1&    &    \hspace{0.5cm} -3.4&     \\ 
$2\times 10^9$ &   14000&	 &    3.3&    &    \hspace{0.5cm} -1.2&    &    \hspace{0.5cm} -5.0&    &    5600&	&	 3.2&    &	 \hspace{0.5cm} -1.2&    &    \hspace{0.5cm} -4.4&     \\ 
$5\times 10^9$ &   19000&	 &    3.4&    &    \hspace{0.5cm} -1.4&    &    \hspace{0.5cm} -6.8&    &	7600&	 &    3.4&    &    \hspace{0.5cm} -1.3&    &    \hspace{0.5cm} -6.1&     \\ 
$1\times 10^{10}$ & 24000&    &    3.6&    &    \hspace{0.5cm} -1.5&    &	\hspace{0.5cm} -8.5&    &	 9600&    &    3.5&    &    \hspace{0.5cm} -1.4&    &	\hspace{0.5cm} -7.6&     \\ 
$2\times 10^{10}$ & 30000&    &    3.7&    &    \hspace{0.5cm} -1.6&    &	\hspace{0.5cm} -10.3&    &    12000&	&	 3.6&    &	 \hspace{0.5cm} -1.5&    &    \hspace{0.5cm} -9.3&     \\ 
$5\times 10^{10}$ & 39000&    &    3.8&    &    \hspace{0.5cm} -1.7&    &	\hspace{0.5cm} -13.1&    &    16000&	&	 3.8&    &	 \hspace{0.5cm} -1.7&    &    \hspace{0.5cm} -11.8&     \\ 
$1\times 10^{11}$ & 48000&    &    4.0&    &    \hspace{0.5cm} -1.8&    &    \hspace{0.5cm} -15.2&    &	 20000&    &	3.9&    &	\hspace{0.5cm} -1.8&    &	 \hspace{0.5cm} -13.9&     \\ 
$2\times 10^{11}$ & 59000&    &    4.1&    &    \hspace{0.5cm} -1.9&    &	\hspace{0.5cm} -17.5&    &    24000&	&	 4.0&    &	\hspace{0.5cm} -1.9&    &	 \hspace{0.5cm} -16.1&     \\ 
$5\times 10^{11}$ & 75000&    &    4.2&    &    \hspace{0.5cm} -2.0&    &    \hspace{0.5cm} -20.3&    &	 31000&    &	4.2&    &	\hspace{0.5cm} -2.0&    &	\hspace{0.5cm} -18.8&     \\ 
\hline
\end{tabular}
\caption{Antineutrino CC and NC cross-sections on isoscalar targets,  
along with their uncertainties, in the perturbative DGLAP formalism at  
NLO, using HERAPDF1.5 (both with and without member 9).}
\label{tbl:nubarHERA}
\end{table}

\end{document}